\documentclass{comnet}

 \usepackage{hyperref} 
\usepackage{graphics,epsfig,graphicx} 

\usepackage{xcolor,url}

\definecolor{vblue}{rgb}{0.19,0.20,0.56}

\hypersetup{
    bookmarks=true,         
    unicode=false,          
    pdftoolbar=true,        
    pdfmenubar=true,        
    pdffitwindow=false,     
    pdfstartview={FitH},    
    pdftitle={My title},    
    pdfauthor={Author},     
    pdfsubject={Subject},   
    pdfcreator={Creator},   
    pdfproducer={Producer}, 
    pdfkeywords={keyword1, key2, key3}, 
    pdfnewwindow=true,      
    colorlinks=true,       
    linkcolor=vblue,          
    citecolor=vblue,        
    filecolor=vblue,      
    urlcolor=vblue          
}

\AtBeginDocument{
  \label{CorrectFirstPageLabel}
  \def\fpage{\pageref{CorrectFirstPageLabel}}
}

\begin{document}

\title{Node Diversification in Complex Networks by Decentralized Coloring}

\shorttitle{Insert short title here for recto running head} 
\shortauthorlist{Garcia-Lebron, Myers, Xu, Sun} 

\author{
\name{Richard Garcia-Lebron}
\address{Department of Computer Science, University of Texas at San Antonio, San Antonio, TX, USA}
\name{David J. Myers}
\address{Air Force Research Laboratory, Rome, NY, USA}
\name{Shouhuai Xu}
\address{Department of Computer Science, University of Texas at San Antonio, San Antonio, TX, USA}
\and
\name{Jie Sun$^\dag$}
\address{Department of Mathematics, Clarkson University, Potsdam, NY, USA and Department of Mechanical \& Aerospace Engineering, Clarkson University, Potsdam, NY, USA
}\email{$^\dag$Corresponding author: sunj@clarkson.edu}}

\maketitle

\begin{abstract}
{We develop a decentralized coloring approach to diversify the nodes in a complex network. The key is the introduction of a local conflict index that measures the color conflicts arising at each node which can be efficiently computed using only local information.  We demonstrate via both synthetic and real-world networks that the proposed approach significantly outperforms random coloring as measured by the size of the largest color-induced connected component. Interestingly, for scale-free networks further improvement of diversity can be achieved by tuning a degree-biasing weighting parameter in the local conflict index.}
{decentralized coloring, network connectivity, iterative optimization}
\end{abstract}

\section{Introduction}
Complex networks have been serving as structural models for many complex systems in the real world, including social networks as well as engineering and biological systems~\cite{NewmanReview2003}.
Among the many properties of complex networks, connectivity and its determining factors have received considerable attention in recent years~\cite{Dorogovtsev2008RMP,Estrada2011book}, especially the role played by connectivity in dynamical processes such as synchronization~\cite{Pecora1998PRL,ArenasSync2008},
information communication~\cite{AlbertNature00}, and epidemic spreading and control~\cite{Pastor-SatorrasPRL2001,Cohen2003PRL}.
Enhanced connectivity is desirable from the point of view of network functioning, such as communication, information sharing, and transfer of energy.
However, connectivity can be a double-edged sword as it is intimately related to the spreading of virus and disease~\cite{Billings2002PLA,Cohen2003PRL,Newman2002PRE,Pastor-SatorrasPRL2001,Pastor-SatorrasPRE2001,WangSRDS03,VanMieghemIEEEACMTON09,XuTAAS2012,zheng2015active}.
For computer networks, {including the peer-to-peer networks that serve as the networking infrastructure of the widely used blockchains in general and cryptocurrency in particular \cite{Nakamoto_bitcoin:a}}, the adverse effect is rooted in the monoculture of computer software systems:
When all computers run the same software operating system (e.g., Microsoft Windows), a single software vulnerability can cause many computers to be compromised especially when the underlying network is highly connected~\cite{Geer2003,Stamp2004}.
For cyber security applications, it has been advocated to overcome this monoculture problem by introducing software {\it diversification}~\cite{Avizienis1985,DBLP:conf/hotsos/ChenCX18a}.
{For complex networks representing human social contacts, coping with the spreading of epidemic diseases demands the use of vaccines, which are often scarce especially when the diseases are new variants of viruses, to immunize few people in the network while maximize the suppression effect, which is fundametally related to the problem of disrupting network clusters.}
The concept of diversification is also important in socioeconomics, given a strong link between structural network diversity and economic development~\cite{Eagle2010Science}.

In this paper we focus on a particular type of network diversification problem, which is closely related to the problem of {\it graph coloring}.
A network of coupled units can be abstracted as a graph $G=(V,E)$ where a pair of nodes $(u,v)\in E$ if and only if $u$ and $v$ are connected by an edge in the graph. A {\it coloring} of the graph is an assignment function {from a set $V$ of nodes to a set of $q$ colors},
\begin{equation}
{	C:V\rightarrow\{1,2,\dots,q\},}
\end{equation}
where $C(u)$ denotes the color assigned to node $u$.
Upon coloring, an edge $(u,v)\in E$ is called {\it defective} if $C(u)=C(v)$, that is, if nodes $u$ and $v$ have the same color. Unlike non-defective edges, the defective edges are assumed to be the transmission channels for the spread of undesired information such as computer virus. {In addition to the problem of coloring planar graphs (also known as the ``map coloring problem" for which the {\it four color theorem} is about),}  the graph coloring abstraction {has several applications, for example in task scheduling problems (including computer register allocation)~\cite{Marx2004}, frequency assignment and planning in wireless communication systems~\cite{Eisenblatter2002}, and more recently in studying the virus propagation problems in various} computer networks~\cite{ODonnell2004,Kuhn2009,DBLP:conf/ccs/XuLP08,DBLP:journals/tifs/XuLPW11,DBLP:conf/hotsos/ChenCX18a,DBLP:journals/tnse/ZhengLX18}.
In these applications, the goal of coloring is to diversify the colors of the nodes so that the network is ``disrupted" into as small as possible {\it independent sets}---sets of isolated components made up of defective edges.

{In this paper we} develop a decentralized coloring approach based on iterative minimization of a {\it local conflict index} (LCI), which is a quantity that is directly computable from the local information at each node. We validate the effectiveness of our method for both random and real-world networks, and found that for scale-free networks further improvement can be achieved by using a degree-biasing weighting scheme.

\section{Decentralized Network Coloring}
For graph $G$ and color assignment $C$, we define the LCI at node $u$ as
\begin{equation}\label{eq:LCI}
	\mbox{LCI}(u)=\sum_{v\in\mathcal{N}_u}w_{v}\delta(C(u),C(v)),
\end{equation}
where $\mathcal{N}_u$ denotes the set of neighbors of $u$ in $G$, and $\delta(i,j)=1$ if $i=j$ and $0$ otherwise.
The weight $w_v$ can be used to adjust the relative contribution of a defective edge $(u,v)$ to $\mbox{LCI}(u)$.

\begin{figure}[htbp]
\centering
\includegraphics*[width=0.65\textwidth]{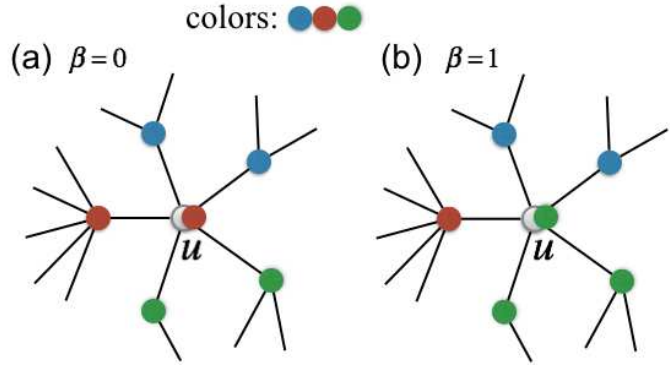}
\caption{Minimization of LCI at a given node $u$ depends on the choice of $\beta$ in Eq.~\eqref{eq:LCI}. As examples: (a) When $\beta=0$, $\mbox{LCI}(u)$ is minimized by choosing a color for $u$ that results in the smallest number of defective edges that connect to $u$. (b) When $\beta=1$, $\mbox{LCI}(u)$ is minimized by choosing a color for $u$ so that the total degree of the defective neighbors of $u$ is minimized.}
\label{fig_cartoon}
\end{figure}

As a first attempt at exploring the effect of the weight, we consider weights of the form
\begin{equation}\label{eq:beta}
	w_v = k_v^\beta,
\end{equation}
where $k_v=|\mathcal{N}_v|$ denotes the degree of node $v$. This weighting scheme allows one to tune the relative influence of node degree on LCI by adjusting exponent $\beta$.
For the special case $\beta=0$, $\mbox{LCI}(u)$ gives the number of defective edges at node $u$ since all weights $w_u=1$.
This is the scenario considered in classical statistical physics such as the Potts model~\cite{Wu1982}.
For any node with degree $k>1$: when $\beta\rightarrow-\infty$, $\mbox{LCI}\rightarrow0$ regardless of the particular color assignment; on the other hand, when $\beta\rightarrow+\infty$, the value of $\mbox{LCI}(u)$ is dominated by the maximum degree of the neighbors of $u$ that have the same color as $u$'s.
When $\beta>0$ ($\beta<0$) and proper coloring at node $u$ is not possible, the minimization of $\mbox{LCI}(u)$ is generally achieved by biasing the coloring of $u$ toward the low-degree (high-degree) neighbors of $u$ that have the same color as $u$'s. An  illustration is provided in Fig.~\ref{fig_cartoon}, which visually highlights the following:
when $\beta=0$, $\mbox{LCI}(u)$ is minimized by choosing a color that results in the least number of defective edges that connect to $u$; when $\beta=1$, $\mbox{LCI}(u)$ is minimized when the total degree of the {\em defective neighbors} of $u$ (i.e., the neighbors connected to $u$ over defective edges) is minimized.
A key feature of $\mbox{LCI}$ is that it can be computed without global knowledge of the network. Importantly, the determination of $\mbox{LCI}(u)$ only involves information about
the colors as well as the degrees of $u$ and its neighbors.

Given a network $G$, a fixed number of colors $q$ and the choice of weights $w_v$, the iterative minimization of $\mbox{LCI}$ over the nodes of a network gives rise to a dynamic coloring process, which we call {\it dynamic decentralized coloring} (DDC). Initially, each node is assigned a color chosen uniformly at random from the $q$ colors (nodes with the same colors form a (site) percolation graph with ``occupation" probability $1/q$~\cite{Moore:2000}.)
Then, in each iteration, a node $u$ is selected randomly, and its color will be updated to minimize $\mbox{LCI}(u)$ given information about the colors and degrees of the neighbors of $u$. When there is no defective edge associated to node $u$, the color of $u$ remains unchanged. On the other hand, when there is at least one defective edge, $u$ updates its color to minimize $\mbox{LCI}(u)$. When multiple color choices yield the same minimal $\mbox{LCI}(u)$, the new color of $u$ will be chosen randomly among these minimizers.
Empirically we found the algorithm converges in $O(ng(n))$ iterations in terms of the fraction of defection edges, that is, $O(g(n))$ color updates per node. The function $g(n)$ generally decreases as the number of colors $q$ increases, from linear (when $q<\chi(G)$) to logarithmic (when $q\gtrsim\chi(G)$) for various types of networks.

\section{Results}
\subsection{{{Measures of diversity}}}
In order to quantify the network diversity upon coloring, we consider the following two measures.
One measure is the fraction of defective edges, given by
\begin{equation}\label{eq:fd}
	f_d =\mbox{number of defective edges}/\mbox{total number of edges}.
\end{equation}
The case of $f_d\rightarrow0$ occurs when there is no defective edge whereas $f_d\rightarrow1$ if and only if all edges are defective, which, for a connected network, can only occur if all nodes have the same color. In general $0\leq f_d\leq 1$, with a smaller value corresponding to a better diversity.
However, $f_d$, as a local measure, has some undesirable limitations.
For example, the coloring may yield a large {\em color-induced component} even if $f_d\approx0$.
We therefore also consider a global diversity measure
\begin{equation}\label{eq:Rmax}
	R_{\max} = \mbox{size of the largest color-induced component}.
\end{equation}
For a given coloring, $R_{\max}$ can be interpreted as the {\it maximum range of spread} as it is the maximum number of nodes that can be reached from a single node through defective edges.

\subsection{{{Random networks}}}
\begin{figure}[htbp]
\centering
\includegraphics*[width=0.75\textwidth]{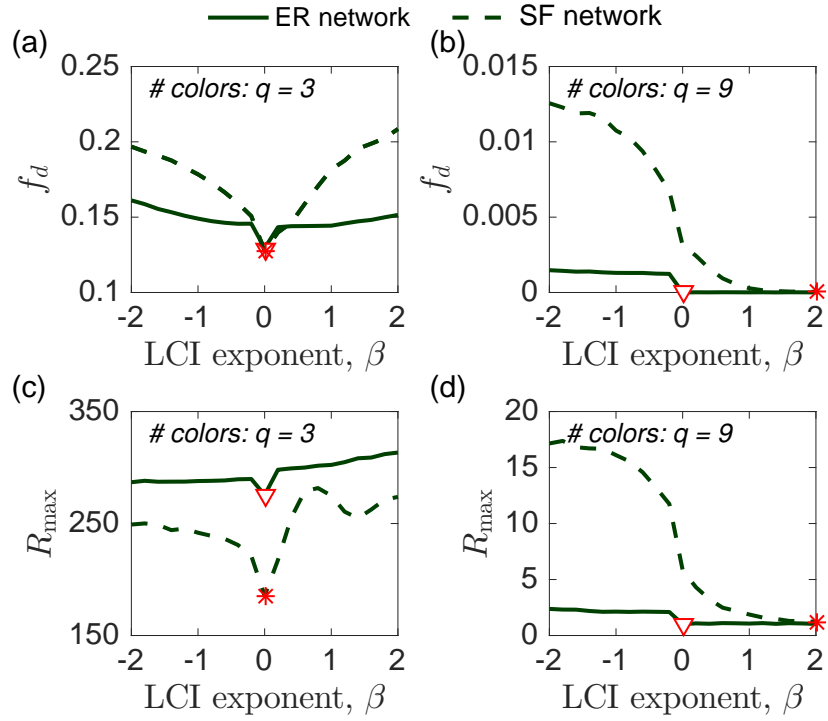}
\caption{
Dependence of network diversity on the choice of $\beta$ in the LCI minimization. (a-b) Fraction of defective edges $f_d$ [as defined in Eq.~\eqref{eq:fd}] as a function of $\beta$ for: ER network with $n=1,000$ nodes and edge probability $p=0.015$ (solid curves), and SF network with $n=1,000$ nodes and degree distribution $P(k)\sim k^{-2.5}$ with minimal degree $k_{\min}=5$ (dashed curves). Locations of the optimal $\beta$ are highlighted by ``$\triangledown$" for the ER network and by ``$\ast$" for the SF network, for $q=3$ colors (a) and $q=9$ colors (b), respectively.
(c-d) Same as (a-b) by considering the maximum range of damage $R_{\max}$ which is a global disruptiveness measure given by Eq.~\eqref{eq:Rmax}.
In all panels, each data point represents the average over $150$ independent runs under the same parameters.}
\label{fig_beta}
\end{figure}

\begin{figure}[htbp]
\centering
\includegraphics*[width=0.75\textwidth]{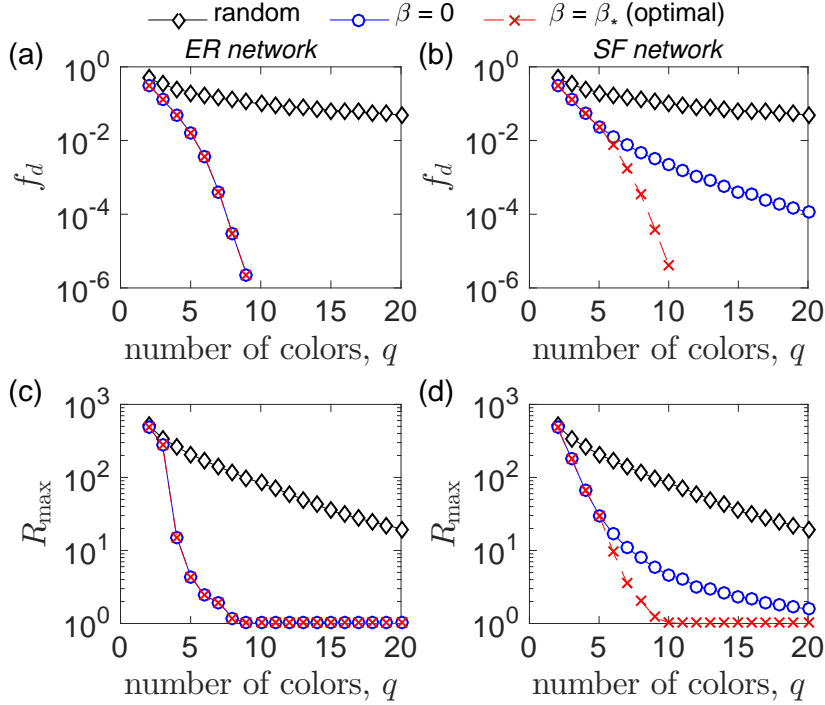}
\caption{
Dependence of network diversity on the number of colors in the three coloring schemes: randomly selecting a color for each node (random), iterative minimization of the LCI [see Eq.~\eqref{eq:LCI}] using $\beta=0$ and the optimal value of $\beta$. (a) Fraction of defective edges $f_d$ [as defined in Eq.~\eqref{eq:fd}] as functions of the number of colors for ER network with $n=1,000$ nodes and edge probability $p=0.015$. (b) Same as (a), for SF network with $n=1,000$ nodes and degree distribution $P(k)\sim k^{-2.5}$ with minimal degree $k_{\min}=5$. (c-d) Same as (a-b) by considering the maximum range of damage $R_{\max}$ given by Eq.~\eqref{eq:Rmax}.
In all panels, each data point represents the average over $150$ independent runs under the same parameters.
}
\label{fig_color}
\end{figure}

We first explore the global effect of the parameter $\beta$ on the coloring of networks. We consider two networks with the same number of nodes ($n=1,000$).
The first network is generated by the classical Erd\H{o}s-R\'{e}nyi (ER) model,
where  an edge is created between every pair of nodes $(u,v)$ with probability $p$~\cite{Erdos60}. Here we choose $p=0.015$, resulting in a sparse network with average degree $\langle k\rangle\approx np=15$.
The second network is a scale-free (SF) network generated by the configuration model~\cite{Bender1978,Bollobas1980} with expected degree distribution $P(k)\approx k^{-\gamma}$. Here we set the degree exponent $\gamma=2.5$ and minimal degree $k_{\min}=5$, obtaining a network with average degree $\langle k\rangle\approx12$.
Both networks have a single connected component. Upon coloring, the networks are expected to be disrupted into same-color components as the number of available colors increases.
Figure~\ref{fig_beta} shows that for the ER network with a fixed number of colors, both the fraction of defective edges $f_d$ [Fig.~\ref{fig_beta}(a,b)] and the maximum range of damage $R_{\max}$ [Fig.~\ref{fig_beta}(c,d)] are minimized when $\beta\approx0$.
Interestingly, the same does not hold true for the SF network. The optimal value of $\beta$ for the SF network in fact depends on the number of colors. When there are very few colors available, the optimal $\beta\approx0$ [Fig.~\ref{fig_beta}(a,c)]; on the other hand, with increasing number of colors, the optimal value of $\beta\gg0$ [Fig.~\ref{fig_beta}(b,d)].
The fact that $f_d$ (the fraction of defective edges in the network) is minimized for a nonzero $\beta$ suggests a nontrivial global effect of the minimization of LCI, because locally the $\mbox{LCI}$ value indeed equals the number of defective edges with the choice of $\beta=0$ (equivalent to greedily minimizing $f_d$ in each iteration).
Next, we study how the number of colors affects network diversity. We consider both the ER network and the SF network as described in Fig.~\ref{fig_beta}, and compute the fraction of defective edges $f_d$ as well as the maximum range of damage $R_{\max}$ as functions of the number of colors. Results from three distinct algorithms are compared: (1) random coloring by choosing a color for each node uniformly at random from the $q$ colors; (2) iterative LCI minimization with $\beta=0$; and (3) iterative LCI minimization with $\beta=\beta_*$, which is the optimal value of $\beta$ for the given number of colors found numerically by searching over $\beta\in(-2,2)$.
The results are shown in Fig.~\ref{fig_color}. In comparison to random coloring, minimization of LCI leads to substantially faster decay of both $f_d$ and $R_{\max}$ as function of $q$, both of which reach the saturation level with $q=10$ colors, at which point a proper coloring is essentially achieved. For the ER network, there is no significant difference whether $\beta$ is chosen to be $0$ or the actual optimal value (which can in fact differ from $\beta$), suggesting that $\beta=0$ effectively leads to optimal diversity  independent of the number of colors $q$. This is consistent with the results shown in Fig.~\ref{fig_beta}(a,c) for ER network,
However, for the SF network, it is in fact possible to achieve significant improvement of network diversity by optimally choosing $\beta$, and such choice is not universal as the ER case but rather depends on the network structure as well as the number of colors available.

\pagebreak

\subsection{{{Networks with two communities}}}
\begin{figure}[htbp]
\centering
\includegraphics*[width=0.75\textwidth]{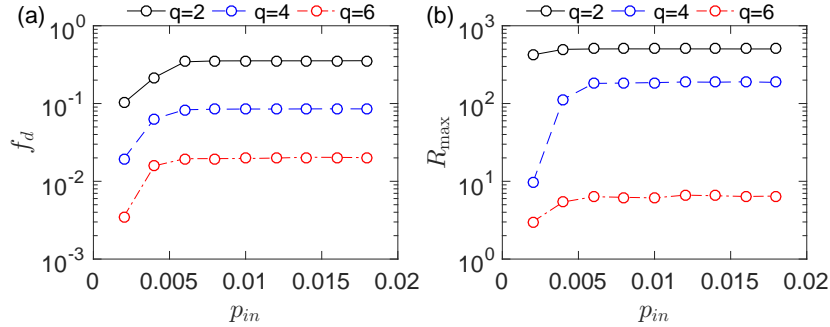}
\caption{{Impact of community structure on coloring. We apply dynamic decentralized coloring to random networks with $n=1,000$ nodes for several different number of colors, $q$. We found that the parameter $\beta$ has the minimal impact on the outcome, and thus fix $\beta=0$ throughout the experiments. For each network, the nodes are divided into two equal-size groups (communities): nodes within the same community are connected with probability $p_{in}$ and nodes from different communities are connected with probability $p_{out}$. To compare networks with different strength of community structure, we vary $p_{in}$ (within-community connection probability) while fixing $p_{in}+p_{out}=0.02$ so that the average degree $\langle k\rangle\approx\frac{n}{2}(p_{in}+p_{out})=10$. We found that network diversification, as measured by $f_d$ and $R_{\max}$, are both worsened as $p_{in}$ increases, that is, as community structure becomes more pronounced.
}}
\label{fig_community}
\end{figure}

{{The types of networks we considered so far are {\it unstructured}, with no community or other structures. Are networks with community structures easier or harder to diversify? As a first attempt to address this question, we consider a random network model with $n$ nodes that are split into two groups. Within each group, two nodes are connected with probability $p_{in}$; two nodes from different groups are connected with probability $p_{out}$. That is, $p_{in}$ and $p_{out}$ are the within-group (or within-community) and cross-group (cross-community) connection probabilities, respectively, The larger the value of $p_{in}$ when compared with $p_{out}$, the stronger the community structure, whereas in the case of $p_{in}=p_{out}$ the model reduces to a standard ER model with no community structure. Through numerical simulations, we found that for networks generated by this model, the outcome of decentralized coloring does not depend much on the parameter $\beta$, similar to the case of Erd\H{o}s-R\'{e}nyi (ER) networks. However, diversity does depend on how strong the communities are, as controlled by the parameters $p_{in}$ and $p_{out}$. Generally speaking, for networks with similar total numbers of edges, networks with a smaller within-community connection probability $p_{in}$ seem easier to diversify (see Fig.~\ref{fig_community}). This also suggests that the presence of community structures makes it harder to diversify the network.}}

\subsection{{{Application: decentralized coloring of an Email communication network}}}
\begin{figure}[htbp]
\centering
\includegraphics*[width=0.75\textwidth]{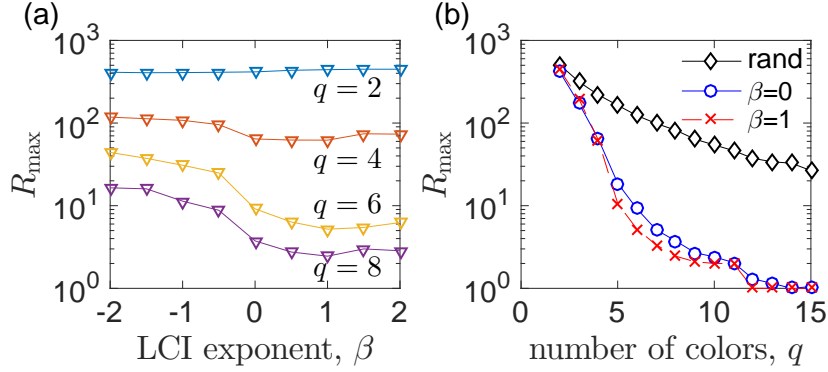}
\caption{Dependence of network diversity as measured by $R_{\max}$ as functions of
(a) the LCI exponent $\beta$ used in the proposed decentralized coloring under different numbers of colors $q$; and
(b) the number of colors under random coloring (``rand") and the proposed decentralized coloring with $\beta=0$ and $\beta=1$, respectively.}
\label{fig_email}
\end{figure}

Finally, noting that email communications represent a primary source of computer virus spreading and the difficulty of managing the system in a centralized fashion, we test our decentralized coloring approach on a real-world email network.
The network we consider was constructed from email communications between members of the {\it University Rovira i Virgili}~\cite{Guimera2003}. We focus on the largest connected component of the network which contains $n=1,133$ nodes and $m=5,451$ edges. The average degree $\langle k\rangle =2m/n\approx9.62$ and the maximum degree $k_{\max}=71$, with the degree distribution reasonably resembled by an exponential, $P(k)\propto \exp(-k/k^*)$ with $k^*\approx9.2$~\cite{Guimera2003}.
In Fig.~\ref{fig_email}(a) we plot $R_{\max}$ as functions of $\beta$ that result from coloring the network via the proposed decentralized algorithm, for varying numbers of colors $q$. As the number of colors increases past $q=4$, the value of $R_{\max}$ tends to be the smallest when $\beta\approx0$. As shown in Fig.~\ref{fig_email}(b), the coloring obtained by the proposed method achieves much better network diversification compared to random coloring, and there is a wide range of number of colors for which the choice of $\beta=1$ outperforms the choice of $\beta=0$.
Given that $R_{\max}$ measures the number of computers that can potentially be infected due to (direct and indirect) communications with a single initially infected computer, these results suggest that our decentralized coloring approach can significantly reduce the risk of large-scale virus outbreak. Such prevention is even more effective by optimizing the weight parameter $\beta$ although such optimization depends intricately on both the network structure and resources available (number of ``colors").

\section{{Conclusion and Discussion}}
In this paper we develop a decentralized coloring approach to diversify the nodes of of a large complex network.
The proposed approach, reminiscent of a self-organized process that involves no central controlling, is achieved by iterative minimization of LCI over the nodes in a network.
In our LCI formulation, the case of weighing parameter $\beta=0$ is equivalent to the antiferromagnetic Potts model at zero temperature~\cite{Wu1982}. By allowing $\beta>0$ which tends to avoid
defective edges connecting to high-degree neighbors, our approach can achieve a significantly improved diversity so long as there is a relatively abundance of colors. Interestingly, such improvement is only observed in scale-free networks but not for Erd\H{o}s-R\'{e}nyi networks which have a rather flat degree distribution. From a theoretical perspective, it remains a challenge to uncover the mechanisms that underlie this discrepancy, which we hope to address in future work. In addition, our coloring algorithm only takes into account the degree of nodes and ignores the effect of clustering. Although the problem is already quite complex and shows rich phenomena even at this relatively simple setting, further incorporation of local clustering information could prove useful for certain types of networks.
{{Furthermore, the preliminary results we reported regarding the dependence of coloring outcome on community structure show that (with fixed number of nodes and edges) networks with more profound community structure are harder to diversify. Interestingly, the {\it graph complement} of such networks have recently been reported to promote diffusion and synchronizability~\cite{Nishikawa2017PRX}. Future research on the relationship between coloring and structural properties of networks will likely yield new insights into how to diversify networks that are more realistic than the rudimentary network models considered here.}}

Among applications, we note that distributed defective coloring algorithms have often been used as an intermediate step for attaining proper coloring \cite{Barenboim2013by,Kuhn2009}. For a network with maximum degree $k_{\max}$, the state-of-the-art distributed defective coloring algorithm \cite{Barenboim2011iw} produces coloring-induced subgraphs whose maximum degree is upper bounded by $k_{\max}/q$.
Empirically we found our algorithm to typically yield a much smaller maximum degree. This hints future research to design decentralized proper-coloring algorithm.
For cybersecurity applications, we tested our coloring approach on an empirical email communication network and found that the network can be effectively ``disrupted" into small disconnected communication components using just a handful of colors. This suggests an opportunity to potentially enhance network communication security via the design and implementation of decentralized software diversification.

{Finally, at a fundamental level, the dynamic decentralized coloring introduced in this paper essentially defines a stochastic process on the network. The coloring of networks is intimately related to the asymptotic behavior of such process. Does the process converge, (if so) does it always converge to the same distribution, how fast, on what types of networks, using how many colors? It is our hope that future work will address these interesting and relevant questions.}

\section*{Acknowledgment}
We thank Daniel ben-Avraham for comments and discussions.
This work was funded in part by the Simons Foundation Grant No. 318812 (J.S.), the Army Research Office Grant No. W911NF-12-1-0276 (J.S.) and No. W911NF-17-1-0566 (R.G. and S.X.), and  NSF grant No. 1736209 (R.G. and S.X.).
This paper was approved for public release; distribution unlimited: 88ABW-2016-0517.


\bibliographystyle{comnet}

\begin{thebibliography}{00}

\bibitem{NewmanReview2003}
Newman, M. (2003)  The structure and function of complex networks. {\em SIAM
  Review}, \textbf{45}, 167.

\bibitem{Dorogovtsev2008RMP}
Dorogovtsev, S.~N., Goltsev, A.~V. {\&} Mendes, J. F.~F. (2008)  Critical
  phenomena in complex networks. {\em Reviews of Modern Physics}, \textbf{80},
  1275.

\bibitem{Estrada2011book}
Estrada, E. (2011) {\em The Structure of Complex Networks: : Theory and Applications}.
Oxford University Press, Oxford, UK.

\bibitem{ArenasSync2008}
Arenas, A., Diaz-Guilera, A., Kurths, J., Moreno, Y. {\&} Zhou, C. (2008)
  Synchronization in complex networks. {\em Phys. Rep.}, \textbf{469},
  93--153.

\bibitem{Pecora1998PRL}
Pecora, L.~M. {\&} Carroll, T.~L. (1998)  Master stability functions for
  synchronized coupled systems. {\em Phys. Rev. Lett.}, \textbf{80}, 2109.

\bibitem{AlbertNature00}
Albert, R., Jeong, H. {\&} Barabasi, A. (2000)  Error and attack tolerance of
  complex networks. {\em Nature}, \textbf{406}, 378--482.

\bibitem{Cohen2003PRL}
Cohen, R., Havlin, S. {\&} ben Avraham, D. (2003)  Efficient Immunization
  Strategies for Computer Networks and Populations. {\em Phys. Rev. Lett.},
  \textbf{91}, 247901.


\bibitem{Pastor-SatorrasPRL2001}
Pastor-Satorras, R. {\&} Vespignani, A. (2001b)  Epidemic Spreading in
  Scale-Free Networks. {\em Phys. Rev. Lett.}, \textbf{86}(14), 3200--3203.

\bibitem{Billings2002PLA}
Billings, L., Spears, W.~M. {\&} Schwartz, I.~B. (2002)  A Unified Prediction
  of Computer Virus Spread in Connected Networks. {\em Phys. Lett. A},
  \textbf{297}, 261--266.

\bibitem{Newman2002PRE}
Newman, M. E.~J., Forrest, S. {\&} Balthrop, J. (2002)  Email networks and the
  spread of computer viruses. {\em Phys. Rev. E}, \textbf{66}, 035101.


\bibitem{Pastor-SatorrasPRE2001}
Pastor-Satorras, R. {\&} Vespignani, A. (2001a)  Epidemic dynamics and endemic
  states in complex networks. {\em Phys. Rev. E}, \textbf{63}, 066117.

\bibitem{VanMieghemIEEEACMTON09}
Van~Mieghem, P., Omic, J. {\&} Kooij, R. (2009)  Virus spread in networks. {\em
  IEEE/ACM Trans. Netw.}, \textbf{17}, 1--14.


\bibitem{WangSRDS03}
Wang, Y., Chakrabarti, D., Wang, C. {\&} Faloutsos, C. (2003)  Epidemic
  Spreading in Real Networks: An Eigenvalue Viewpoint. In {\em Proc. of the
  22nd IEEE Symposium on Reliable Distributed Systems (SRDS'03)}, pp. 25--34.
  \url{https://ieeexplore.ieee.org/document/1238052}

\bibitem{XuTAAS2012}
Xu, S., Lu, W. {\&} Xu, L. (2012)  Push- and pull-based epidemic spreading in
  networks: Thresholds and deeper insights. {\em ACM Trans. Auton. Adap. Syst. (ACM TAAS)}, \textbf{7}, 32. \url{ https://dl.acm.org/citation.cfm?id=2746196}

\bibitem{zheng2015active}
Zheng, R., Lu, W. {\&} Xu, S. (2015)  Active cyber defense dynamics exhibiting
  rich phenomena. In {\em Proceedings of the 2015 Symposium and Bootcamp on the
  Science of Security}. ACM, p. 2.

\bibitem{Nakamoto_bitcoin:a}
Nakamoto, S. (2009)  Bitcoin: A peer-to-peer electronic cash system.
  \url{http://bitcoin.org/bitcoin.pdf}. Last access: April 20, 2018.

\bibitem{Geer2003}
Geer, D., Bace, R., Gutmann, P., Metzger, P., Pfleeger, C.~P., Quarterman,
  J.~S. {\&} Schneier, B. (27 September 2003)  CyberInsecurity: The Cost of
  Monopoly. \url{http://cryptome.org/cyberinsecurity.htm}. Last access: April 20, 2018.

\bibitem{Stamp2004}
Stamp, M. (2004)  Risks of Monoculture. {\em Commun. ACM}, \textbf{47},
  120.

\bibitem{Avizienis1985}
Avizienis, A. (1985)  The N-Version Approach to Fault-Tolerant Software. {\em
  IEEE Trans. Softw. Eng.}, \textbf{11}, 1491--1501.

\bibitem{DBLP:conf/hotsos/ChenCX18a}
Chen, H., Cho, J. {\&} Xu, S. (2018)  Quantifying the security effectiveness of
  network diversity: poster. In {\em Proceedings of the 5th Annual Symposium
  and Bootcamp on Hot Topics in the Science of Security (HoTSoS'2018)}, pp.
  24:1.
\url{https://dl.acm.org/citation.cfm?doid=3190619.3191680}

\bibitem{Eagle2010Science}
Eagle, N., Macy, M. {\&} Claxton, R. (2010)  Network Diversity and Economic
  Development. {\em Science}, \textbf{328}, 1029--1031.

\bibitem{Marx2004}
Marx, D. Graph colouring problems and their applications in scheduling. {\em
  Periodica Polytechnica Electrical Engineering}, \textbf{48}, 11--16.

\bibitem{Eisenblatter2002}
Eisenbl{\"a}tter, A., Gr{\"o}tschel, M. {\&} Koster, A. M. C.~A. (2002)
  Frequency planning and ramifications of coloring. {\em Discussiones
  Mathematicae Graph Theory}, \textbf{22}, 51--88.

\bibitem{Kuhn2009}
Kuhn, F. (2009)  Weak Graph Colorings: Distributed Algorithms and Applications.
  In {\em Proceedings of the Twenty-first Annual Symposium on Parallelism in
  Algorithms and Architectures}, SPAA '09, pp. 138--144, New York, NY, USA.
  ACM. \url{https://dl.acm.org/citation.cfm?doid=1583991.1584032}

  \bibitem{ODonnell2004}
O'Donnell, A.~J. {\&} Sethu, H. (2004)  On Achieving Software Diversity for
  Improved Network Security Using Distributed Coloring Algorithms. In {\em
  Proceedings of the 11th ACM Conference on Computer and Communications
  Security}, CCS '04, pp. 121--131, New York, NY, USA. ACM.
  \url{https://dl.acm.org/citation.cfm?doid=1030083.1030101}

  \bibitem{DBLP:conf/ccs/XuLP08}
Xu, S., Li, X. {\&} Parker, T.~P. (2008)  Exploiting social networks for
  threshold signing: attack-resilience vs. availability. In {\em Proceedings of
  the 2008 {ACM} Symposium on Information, Computer and Communications
  Security, {ASIACCS} 2008, Tokyo, Japan, March 18-20, 2008}, pp. 325--336.
  \url{https://dl.acm.org/citation.cfm?doid=1368310.1368358}

  \bibitem{DBLP:journals/tifs/XuLPW11}
Xu, S., Li, X., Parker, T.~P. {\&} Wang, X. (2011)  Exploiting Trust-Based
  Social Networks for Distributed Protection of Sensitive Data. {\em {IEEE}
  Trans. Information Forensics and Security}, \textbf{6}, 39--52.

 \bibitem{DBLP:journals/tnse/ZhengLX18}
Zheng, R., Lu, W. {\&} Xu, S. (2018)  Preventive and Reactive Cyber Defense
  Dynamics Is Globally Stable. {\em {IEEE} Trans. Network Science and
  Engineering}, \textbf{5}, 156--170.

  \bibitem{Wu1982}
Wu, F.~Y. (1982)  The Potts Model. {\em Rev. Mod. Phys.}, \textbf{54}, 235.


\bibitem{Moore:2000}
{Moore}, C. {\&} {Newman}, M.~E.~J. (2000)  {Exact solution of site and bond
  percolation on small-world networks}. {\em Physical Review E}, \textbf{62},
  7059--7064.


\bibitem{Erdos60}
Erdos, P. {\&} Renyi, A. (1960)  On the Evolution of random graphs. {\em
  Publications of the Mathematical Institute of the Hungarian Academy of
  Sciences}, \textbf{5}, 17--61.


\bibitem{Bender1978}
Bender, E.~A. {\&} Canfield, E.~R. (1978)  The asymptotic number of labeled
  graphs with given degree sequences. {\em J. Combin. Theor. A}, \textbf{24},
  296.

  \bibitem{Bollobas1980}
Bollob\'{a}s, B. (1980)  A probabilistic proof of an asymptotic formula for the
  number of labelled regular graphs. {\em Eur. J. Eur. J. Comb.}, page {\textbf{1}, 311--316}.

\bibitem{Guimera2003}
Guimera, R., Danon, L., Diaz-Guilera, A., Giralt, F. {\&} Arenas, A. (2003)
  Self-similar community structure in a network of human interactions.. {\em
  Phys. Rev. E}, \textbf{68}, 065103(R).


\bibitem{Nishikawa2017PRX}
Nishikawa, T., Sun, J. {\&} Motter, A.~E. (2017)  Sensitive Dependence of
  Optimal Network Dynamics on Network Structure. {\em Phys. Rev. X},
  \textbf{7}, 041044.


\bibitem{Barenboim2013by}
Barenboim, L. {\&} Elkin, M. (2013)  {Distributed Graph Coloring: Fundamentals
  and Recent Developments}. {\em vol. 4. Morgan {\&} Claypool Publishers,} pp. {1--171}.

\bibitem{Barenboim2011iw}
Barenboim, L. {\&} Elkin, M. (2011)  { Distributed deterministic edge coloring using bounded neighborhood independence}. {\em Distributed Computing}, {\textbf{26}, 273-287}.


































\end{thebibliography}

\begin{thebibliography}{00}

\bibitem{AlbertNature00}
Albert, R., Jeong, H. {\&} Barabasi, A. (2000)  Error and attack tolerance of
  complex networks. {\em Nature}, \textbf{406}, 378--482.

\bibitem{ArenasSync2008}
Arenas, A., Diaz-Guilera, A., Kurths, J., Moreno, Y. {\&} Zhou, C. (2008)
  Synchronization in complex networks. {\em Phys. Rep.}, \textbf{469}(3),
  93--153.

\bibitem{Avizienis1985}
Avizienis, A. (1985)  The N-Version Approach to Fault-Tolerant Software. {\em
  IEEE Trans. Softw. Eng.}, \textbf{11}(12), 1491--1501.

\bibitem{Barenboim2011iw}
Barenboim, L. {\&} Elkin, M. (2011)  {Distributed deterministic edge coloring
  using bounded neighborhood independence.}. {\em PODC}, pages 129--138.

\bibitem{Barenboim2013by}
Barenboim, L. {\&} Elkin, M. (2013)  {Distributed Graph Coloring: Fundamentals
  and Recent Developments}. {\em Morgan {\&} Claypool Publishers 2013},
  \textbf{4}(1), 1--171.

\bibitem{Bender1978}
Bender, E.~A. {\&} Canfield, E.~R. (1978)  The asymptotic number of labeled
  graphs with given degree sequences. {\em J. Combin. Theor. A}, \textbf{24},
  296.

\bibitem{Billings2002PLA}
Billings, L., Spears, W.~M. {\&} Schwartz, I.~B. (2002)  A Unified Prediction
  of Computer Virus Spread in Connected Networks. {\em Phys. Lett. A},
  \textbf{297}, 261--266.

\bibitem{Bollobas1980}
Bollob\'{a}s, B. (1978)  A probabilistic proof of an asymptotic formula for the
  number of labelled regular graphs. {\em Eur. J. Eur. J. Comb.}, page 311.

\bibitem{Cohen2003PRL}
Cohen, R., Havlin, S. {\&} ben Avraham, D. (2003)  Efficient Immunization
  Strategies for Computer Networks and Populations. {\em Phys. Rev. Lett.},
  \textbf{91}, 247901.

\bibitem{Dorogovtsev2008RMP}
Dorogovtsev, S.~N., Goltsev, A.~V. {\&} Mendes, J. F.~F. (2008)  Critical
  phenomena in complex networks. {\em Reviews of Modern Physics}, \textbf{80},
  1275.

\bibitem{Eagle2010Science}
Eagle, N., Macy, M. {\&} Claxton, R. (2010)  Network Diversity and Economic
  Development. {\em Science}, \textbf{328}, 1029--1031.

\bibitem{Eisenblatter2002}
\textcolor{blue}{Eisenbl\"{a}tter, A., Gr\"{o}tschel, M. {\&} Koster, A.M.C.A. (2002)
Frequency planning and ramifications of coloring.
{\em Discussiones Mathematicae Graph Theory} {\bf 22}, 51-88.
}

\bibitem{Erdos60}
Erdos, P. {\&} Renyi, A. (1960)  On the Evolution of random graphs. {\em
  Publications of the Mathematical Institute of the Hungarian Academy of
  Sciences}, \textbf{5}, 17--61.

\bibitem{Estrada2011book}
Estrada, E. (2011) {\em The Structure of Complex Networks}.
Oxford University Press.

\bibitem{Guimera2003}
Guimera, R., Danon, L., Diaz-Guilera, A., Giralt, F. {\&} Arenas, A. (2003)
  {\em Phys. Rev. E}\textbf{68}, 065103(R).

\bibitem{Kuhn2009}
Kuhn, F. (2009)  Weak Graph Colorings: Distributed Algorithms and Applications.
  In {\em Proceedings of the Twenty-first Annual Symposium on Parallelism in
  Algorithms and Architectures}, SPAA '09, pages 138--144, New York, NY, USA.
  ACM.

\bibitem{Marx2004}
\textcolor{blue}{
Marx, D. (2004) Graph colouring problems and their applications in scheduling.
{\em Periodica Polytechnica Ser. El. Eng.} {\bf48}, 11.}

\bibitem{NewmanReview2003}
Newman, M. (2003)  The structure and function of complex networks. {\em SIAM
  Review}, \textbf{45}, 167.

\bibitem{Newman2002PRE}
Newman, M. E.~J., Forrest, S. {\&} Balthrop, J. (2002)  Email networks and the
  spread of computer viruses. {\em Phys. Rev. E}, \textbf{66}, 035101.

\bibitem{Nishikawa2017PRX}
Nishikawa, T., Sun, J., Motter, A.~E. (2017) Sensitive dependence of optimal network dynamics on network structure.
{\em Phys. Rev. X}, \textbf{7}, 041044.

\bibitem{ODonnell2004}
O'Donnell, A.~J. {\&} Sethu, H. (2004)  On Achieving Software Diversity for
  Improved Network Security Using Distributed Coloring Algorithms. In {\em
  Proceedings of the 11th ACM Conference on Computer and Communications
  Security}, CCS '04, pages 121--131, New York, NY, USA. ACM.

\bibitem{Pastor-SatorrasPRE2001}
Pastor-Satorras, R. {\&} Vespignani, A. (2001a)  Epidemic dynamics and endemic
  states in complex networks. {\em Phys. Rev. E}, \textbf{63}(6).

\bibitem{Pastor-SatorrasPRL2001}
Pastor-Satorras, R. {\&} Vespignani, A. (2001b)  Epidemic Spreading in
  Scale-Free Networks. {\em Phys. Rev. Lett.}, \textbf{86}(14), 3200--3203.

\bibitem{Pecora1998PRL}
Pecora, L.~M. {\&} Carroll, T.~L. (1998)  Master stability functions for
  synchronized coupled systems. {\em Phys. Rev. Lett.}, \textbf{80}, 2109.

\bibitem{Stamp2004}
Stamp, M. (2004)  Risks of Monoculture. {\em Commun. ACM}, \textbf{47}(3),
  120--.

\bibitem{Wu1982}
Wu, F.~Y. (1982)  The Potts Model. {\em Rev. Mod. Phys.}, \textbf{54}, 235.

\end{thebibliography}

%







\end{document}